\documentstyle[twoside,fleqn,rgs_workshop,psfig]{article}
\newcommand{\etal}{{\it et al.}}

\newcommand{\asca}{{\small \it ASCA}}

\newcommand{\rrc}{{\small RRC}}
\newcommand{\xmm}{{\it XMM-Newton}}
\newcommand{\rgs}{{\small RGS}}

\newcommand{\ccd}{{\small CCD}}

\newcommand{\chandra}{{\it Chandra}}
\newcommand{\hetgs}{{\small HETGS}}
\newcommand{\hetg}{{\small HETG}}
\newcommand{\heg}{{\small HEG}}

\newcommand{\uv}{{\small UV}}

\newcommand{\nasa}{{\small NASA}}

\newcommand{\cak}{{\small CAK}}
\newcommand{\hmxb}{{\small HMXB}}
\newcommand{\lmxb}{{\small LMXB}}
\newcommand{\vela}{Vela~X-1}
\newcommand{\cen}{Cen~X-3}
\newcommand{\cygxone}{Cyg~X-1}
\newcommand{\cygxthree}{Cyg~X-3}
\newcommand{\gx}{{\small GX}301-2}
\newcommand{\cirxone}{Cir~X-1}
\newcommand{\kalpha}{K$\alpha$}

\newcommand\la{\>\vcenter{\hbox{$<$\hskip-.75em\lower1.0ex\hbox{$\sim$}}}\>}
\newcommand\ga{\>\vcenter{\hbox{$>$\hskip-.75em\lower1.0ex\hbox{$\sim$}}}\>}

\begin{document}

\title{Structure and Dynamics of Stellar Winds in High-mass X-ray Binaries}

\author{
M. Sako,\thanks{{\it Chandra} fellow}\address{\small{Theoretical Astrophysics and
Space Radiation Laboratory, MC 130-33, Caltech, Pasadena, CA 91125,
USA}\vspace{-0.25cm}}
S. M. Kahn,$^{\rm{b}}$ F. Paerels,\address{\small{Columbia Astrophysics
Laboratory, 550 W. 120th St., New York, NY 10027, USA}\vspace{-0.25cm}}
D. A. Liedahl,\address{\small{Physics Department, Lawrence Livermore National
Laboratory, P.O. Box 808, L-41, Livermore, CA 94550, USA}\vspace{-0.25cm}}
S. Watanabe,$^{\rm{de}}$ F. Nagase,$^{\rm{d}}$ \&
T. Takahashi\address{\small{Institute of Space and Astronautical Science 3-1-1
Yoshinodai, Sagamihara, Kanagawa, Japan
229-8510}\vspace{-0.25cm}}\address{\small{Department of Physics, University of
Tokyo, 7-3-1 Hongo, Bunkyo, Tokyo, Japan, 113-0033}\vspace{0.3cm}}
}

\begin{abstract}

  A review of spectroscopic results obtained from \chandra\ High Energy
  Transmission Grating Spectrometer (\hetgs) and \xmm\ Reflection Grating
  Spectrometers (\rgs) observations of several wind-fed high-mass X-ray
  binaries (\hmxb s) is presented.  These observations allow us to study the
  structure of the stellar wind in more detail and provide, for the first
  time, a dyanmical view of the X-ray photoionized wind that surrounds the
  compact object.  At the same time, however, they are also providing us with
  numerous puzzles that cannot be explained in terms of simple models.  For
  example, simple spherically-symmetric wind models cannot explain the
  observed orbital-phase variability of the line intensities and shapes, which
  may be caused by intrinsic asymmetries due to the presence of the compact
  object and/or more complicated radiative transfer effects.  The observed
  line shifts are smaller than those expected from extensions of simple wind
  models of isolated OB supergiants.  In addition, several novel spectroscopic
  discoveries have been made, including: (1) P-Cygni lines from an expanding
  wind, (2) detection of multiple Si K fluorescent lines from a wide range of
  charge states, (3) Compton scattered Fe K lines from a cold medium.  We
  discuss how these spectroscopic diagnostics can be used to understand some
  of the global properties of stellar winds in \hmxb s.

\end{abstract}

\maketitle

\section{Introduction}

  In wind-driven high mass X-ray binaries (\hmxb s), a neutron star or a black
  hole sweeps up a small fraction of a stellar wind lost by a massive O- or
  B-type companion star.  As matter is accreted onto the compact object, a
  fraction of the gravitational potential energy is converted into X-rays,
  which then ionizes and heats the surrounding gas.  The wind reprocesses hard
  X-rays from the compact object, resulting in discrete emission lines and
  continuum radiation that carries a wealth of information about the physical
  state of the reprocessing medium.  The compact object can, therefore, be
  used as an illuminating source to probe the structure of the stellar wind
  and derive the physical parameters that characterize its nature.

  In addition to heating and ionizing the stellar wind, the X-ray source is
  also responsible for dynamically affecting the wind through its intense
  gravitational field and by destroying ions with strong \uv\ resonance
  transitions that drive the stellar outflow.  These effects, in combination
  with the orbital motion of the binary system, also produce a ``wake'' that
  trails behind the compact object.  In low-luminosity \hmxb s, however, the
  degree of stellar wind disruption by the compact object is expected to
  produce only a minor perturbation to the observed X-ray spectrum.  This is
  especially true when observed during eclipse, since many of the dynamical
  effects are produced primarily in the vicinity of the compact object.
  
  \asca ~observations of several \hmxb s have shown that their X-ray spectra
  exhibit both soft X-ray emission from highly ionized ions and fluorescent
  lines from cold, less ionized material (\vela\ -- Nagase \etal\ 1994; \cen\
  -- Ebisawa \etal\ 1996; \gx\ -- Saraswat \etal\ 1996).  Although the
  dominant excitation mechanism (i.e., collisional or photoionization-driven)
  that is responsible for producing the soft X-ray lines cannot be determined
  unambiguously from these observations, cascades following recombination
  seemed to be the most natural candidate given the presence of an intense
  X-ray continuum radiation field.  Subsequently, Liedahl \& Paerels (1996)
  and Kawashima \& Kitamoto (1996), for the first time, detected a narrow
  radiative recombination continuum (\rrc) of S~{\small XVI} in the \asca\
  spectrum of \cygxthree, which provided unequivocal evidence that the lines
  from highly-ionized gas in \cygxthree, and most likely in all \hmxb s, are
  driven almost exclusively through photoionization.

\begin{figure}[t]
  \centerline{\psfig{file=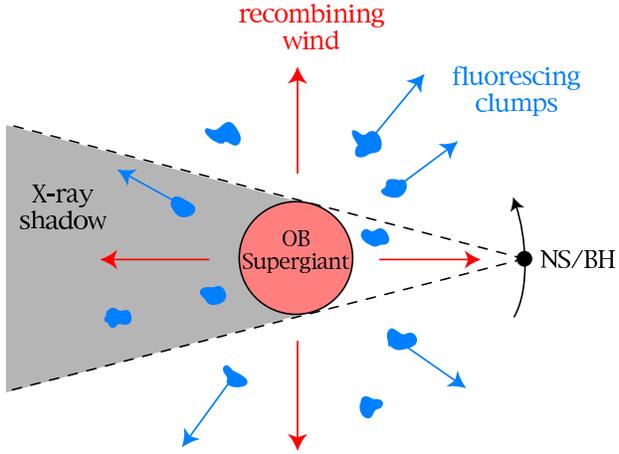,width=3.2in}}

  \caption{A schematic drawing showing the various componets of an
  intermediate-luminosity ($L_X \la 10^{37}~\rm{erg~s}^{-1}$) \hmxb\ system.
  For more luminous systems, the dense clumps may also be highly ionized,
  emitting H- and He-like recombination lines and showing very little
  fluorescence emission. }\label{fig:smallfig}

\end{figure}

  The first physically-motivated modeling of the X-ray spectrum was presented
  by Sako \etal\ (1999) using the same \asca\ spectrum of \vela\ originally
  published by Nagase \etal\ (1994).  Sako \etal\ (1999) have argued that the
  presence of emission lines from both high-ionized and cold near-neutral
  material can be understood only if the wind is structurally inhomogeneous,
  consisting of cool dense clouds embedded in a hot, highly ionized medium.
  They characterized the wind velocity profile as derived by Castor, Abbott,
  \& Klein (1975; hereafter, \cak),
\begin{equation}
  \label{eq:cak}
  v(r) = v_\infty \left(1 - \frac{R_\ast}{r}\right)^\beta,
\end{equation}
  where $R_\ast$ is the radius of the companion star and $v_\infty$ is the
  terminal velocity and $\beta$ is a number that typically lies in the range
  $0.5 - 1.0$ (Abbott 1986; Pauldrach, Puls, \& Kudritzki 1986).  The particle
  density is, then, uniquely specified everywhere in the wind for an assumed
  $v_\infty$, mass loss rate $\dot{M}$, and $\beta$ according to,
\begin{equation}
  \label{eq:density}
  n(r) = \frac{\dot{M}}{4\pi r^2 \mu m_p v_\infty} \left(1 -
  \frac{R_\ast}{r}\right)^{-\beta}.
\end{equation}
  and, hence, the ionization parameter $\xi = L_X/(nr^2)$ is also uniquely
  determined for a given X-ray luminosity.  Here, $r$ is measured from the
  center of the companion star and $\mu$ is the average atomic mass unit of
  the constituent particles ($\mu \approx 1.3$ for solar abundances).  Using
  this relatively simple model for the wind dynamics, they were able to infer
  the velocity structure of the hot medium and the mass-loss rate associated
  with that component.  They also estimated the total mass in the clouds
  through measurements of fluorescent line intensities, and concluded that the
  cold fluorescing material contains a large fraction of the total wind mass
  ($> 90\%$), while most of the wind volume ($> 95\%$) is occupied by highly
  ionized matter.  According to this picture, the stellar wind in Vela X-1 is
  driven by \uv\ radiation from the star that provides a driving force on the
  clumpy material, while the hot, ionized component is essentially transparent
  to \uv\ radiation.  Subsequently, Wojdowski, Liedahl, \& Sako (2001)
  presented a detailed analysis of the X-ray spectrum of \cen, which is on
  average more luminous than \vela\ by roughly an order of magnitude, and
  showed that the dense clumps that produce fluorescence lines in \vela -like
  systems are more highly ionized in \cen.  The mass loss rate inferred from
  H- and He-like ions alone was consistent with those of normal isolated O
  stars.

  While the most general characteristics of stellar winds in \hmxb s were
  revealed using moderate spectral resolution data acquired with \asca, there
  are still many outstanding issues that can only be addressed with high
  resolution spectroscopic data available from \chandra\ and \xmm.  The most
  significant improvement perhaps is the ability to study the dynamics of the
  X-ray emitting gas, which cannot be studied with the \asca\ detectors,
  because line shifts and widths due to typical wind velocities of $\sim
  1000~\rm{km~s}^{-1}$ require a spectral resolving power an order of
  magnitude higher than that of \ccd s ($R$ of at least $\sim 300$).  Accurate
  column density measurements are also possible through measurements of line
  ratios of the He-like triplets and other transitions (Kinkhabwala et
  al. 2002).  In the following sections, we discuss some of the recent
  advances made with observations of \hmxb s with the grating spectrometers on
  \chandra\ and \xmm.

\section{Some General Spectral Properties of Individual Sources}

\begin{figure}[htp]
  \centerline{\psfig{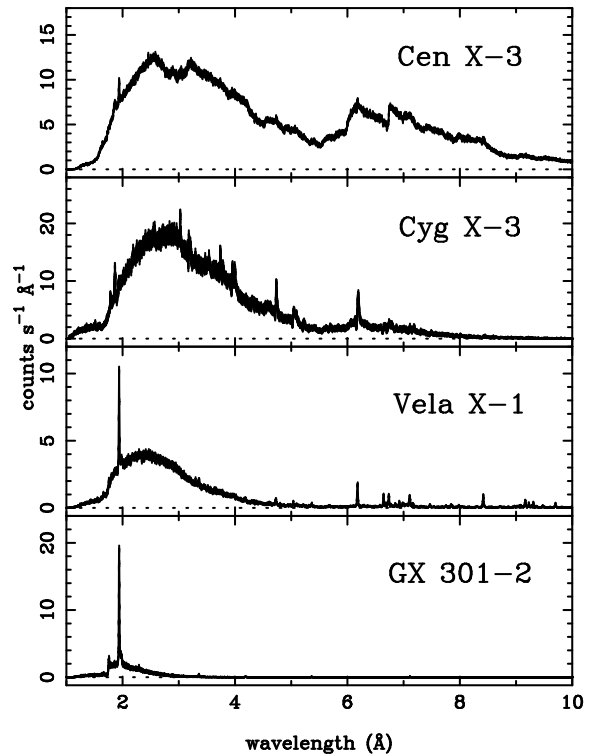}}
  \caption{Global spectral properties of four \hmxb s observed with the
  \chandra\ \hetgs.  Note the differences in the continuum and emission line
  properties. }\label{fig:hmxb_specs}
\end{figure}

  Shown in Figure~\ref{fig:hmxb_specs} are the \chandra\ \hetgs\ spectra of
  four objects \cen, \cygxthree, \vela, and \gx.  As immediately evident from
  the figure, the general properties of both the continuum and line emission
  are vastly different among each of the sources.  The spectrum of \cen\ (1st
  panel in Figure~\ref{fig:hmxb_specs}), which is a relatively high-lumonisity
  system ($L_X \sim 10^{38}~\rm{erg~s}^{-1}$) containing a pulsar with a O 6-8
  {\small III} companion, is completely dominated by continuum emission with a
  relatively low line-of-sight column density of cold material ($N_{\rm{H}}
  \la 10^{22}~\rm{cm}^{-2}$).  The spectrum exhibits very weak absorption
  lines mostly from H-like ions with column densities of $\la
  10^{16}~\rm{cm}^{-2}$.  During eclipse, however, the spectrum is dominated
  almost entirely by line emission as presented by Wojdowski \etal\ (2003).

  In \cygxthree\ (second panel in Figure~\ref{fig:hmxb_specs}), the emission
  line equivalent widths are larger (see, Paerels \etal\ 2000).  This suggests
  that a substantial fraction of the continuum radiation is reprocessed by the
  wind, which implies that the product of the covering fraction and the column
  density of highly ionized material ($\Delta \Omega \times N_i$) is higher.
  Since the X-ray luminosities of \cygxthree\ and of \cen\ during these
  observations were similar, the observed differences in the line spectrum
  also suggest that the average density of the X-ray emission line regions in
  \cygxthree\ is higher.  This is in qualitative agreement with the high
  mass-loss rate of the Wolf-Rayet companion and the compacteness of the
  binary system (van Kerkwijk 1995).

  Local anisotropies in the distribution of absorbing material around the
  X-ray source can also affect the line equivalent widths.  In the \vela\
  spectrum (third panel in Figure~\ref{fig:hmxb_specs}), the observed
  continuum is highly absorbed while the lines longward of $\sim 6$\AA\ have
  high equivalent widths.  This implies that the emission line regions are
  illuminated by a continuum that is less absorbed than what we observe.  In
  other words, the absorber that covers the continuum source along our line of
  sight does not block the emission line regions.  As in \cen, the spectrum of
  \vela\ during eclipse is also dominated by emission lines produced through
  photoionization of the extended stellar wind, as shown by Schulz \etal\
  (2002).

  The soft X-ray photons of \gx\ (last panel in Figure~\ref{fig:hmxb_specs})
  suffers extremely high intrinsic attentuation ($N_H \ga
  10^{23}~\rm{cm}^{-2}$).  The weaknesses of the soft X-ray lines here suggest
  that both the line and continuum emission regions are absorbed, most likely
  by the same material.  Also note the differences in the iron K$\alpha$ line
  equivalents widths.  The more luminous systems (\cen and \cygxthree) with
  $L_X \sim 10^{38}~\rm{erg~s}^{-1}$ show much weaker lines compared to the
  less luminous systems (\vela\ and \gx), which have $L_X \sim
  10^{37}~\rm{erg~s}^{-1}$.

\begin{figure}[t]
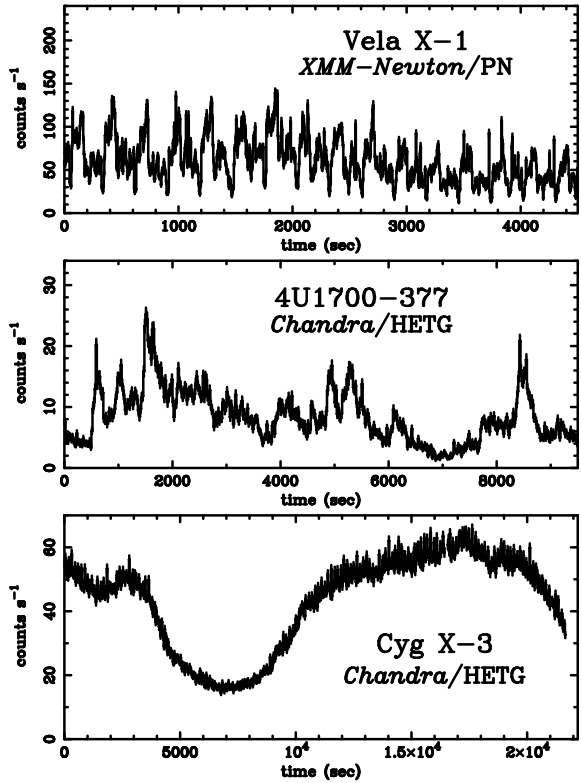

  \centerline{\psfig{file=f3a.ps,width=3.0in,angle=-90}}
  \centerline{\psfig{file=f3b.ps,width=3.0in,angle=-90}}
  \centerline{\psfig{file=f3c.ps,width=3.0in,angle=-90}}
  \caption{X-ray light curves of \vela, an X-ray pulsar with a period of $\sim
  280$ sec (top), 4U1700-377, a neutron star system that shows no pulsations
  (middle), and \cygxthree, a black hole candidate
  (bottom).}\label{fig:hmxb_lcs}
\end{figure}

  The X-ray light curves differ substantially among the various systems as
  well.  Many sources contain pulsars, which show periodic modulations of with
  pulsed fractions of $\sim 50$\% and higher.  These systems also show
  flare-like events during which the luminosity variabilies by more than a
  factor of few on timescales as short as tens of seconds, mostly due to
  accretion instabilites as well as variable absorption on longer timescales
  (see Figure~\ref{fig:hmxb_lcs}).  On the other hand, there are systems like
  \cygxthree, a black hole candidate, which generally exhibits a smooth X-ray
  light curve.  The accretion behavior, therefore, appear to be significantly
  different from source to source.

\section{Wind Dynamics}

  One of the most important improvements made by observations of \hmxb s with
  \chandra\ and \xmm\ is the ability to measure velocities down to a fraction
  of the terminal wind velocity.  In many cases, the observed lines are
  resolved and show complex detailed structure as well as temporal behavior.

\begin{figure}[t]
  \centerline{\psfig{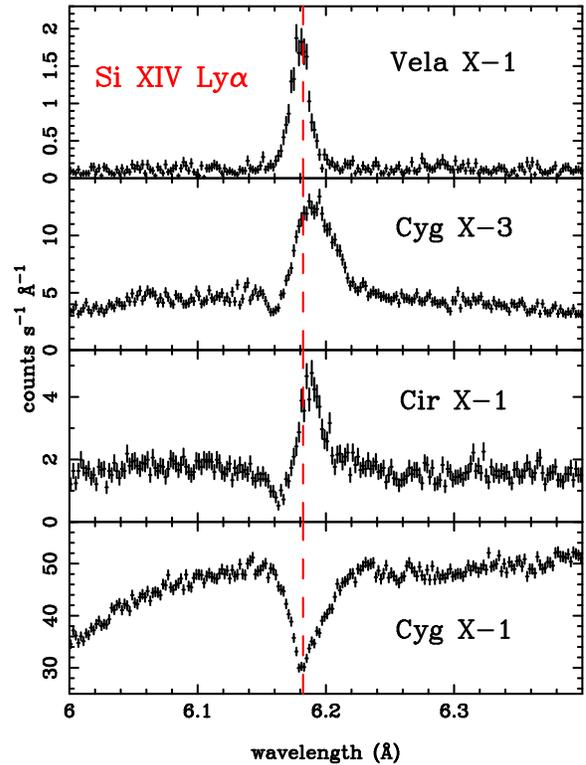}}

  \caption{Observed Si~{\small XIV} line profiles in four different sources
  with varying relative contributions from emission and absorption.  The
  vertical dashed line indicates the rest wavelength of the Si~{\small XIV}
  line. }\label{fig:p-cygni}

\end{figure}

\subsection{Line Profiles and P-Cygni Lines}

\begin{figure}[ht]
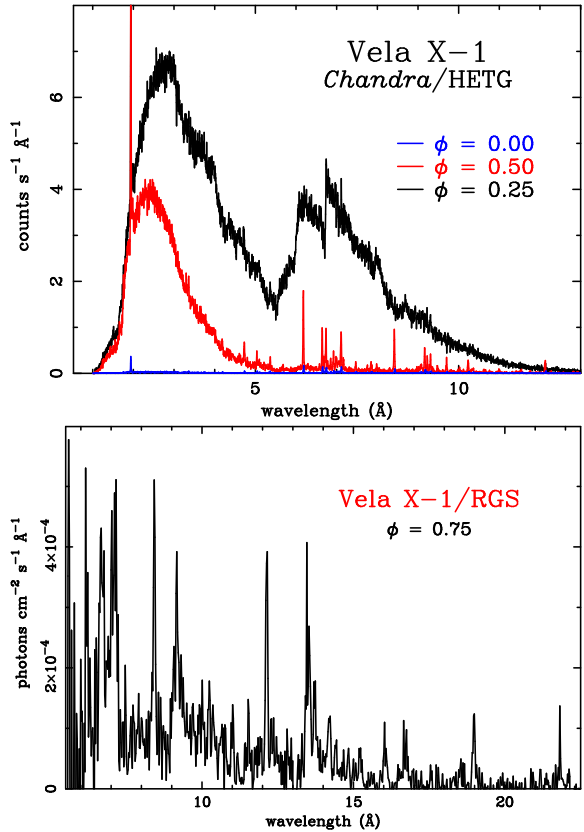

  \centerline{\psfig{file=f5a.ps,width=3.0in,angle=-90}}
  \centerline{\psfig{file=f5b.ps,width=3.0in,angle=-90}}
  \caption{The \chandra\ data of \vela\ observed during three different
  orbital phases (centered at $\phi = 0.0, 0.25$, and $0.5$; top) and the
  \xmm\ spectrum observed at $\phi = 0.75$ (bottom). }\label{fig:vela_phase}
\end{figure}

  The observed line profiles come in many different types as shown in
  Figure~\ref{fig:p-cygni}, and they provide valuable information not only
  about the dynamical properties of the emission line regions, but they are
  also sometimes useful for determining the dominant excitation mechanism as
  well.  The top panel shows a pure emission profile as observed in \vela.  On
  the panel below is a P-Cygni line, which shows a broad emission feature with
  a weak absorption trough towards the blue.  The emission component is much
  stronger than the absorption component, suggesting that recombination is the
  dominant line formation mechanism, since there is not enough continuum
  radiation to account for all the observed line photons via resonant
  fluorescence scattering.  In the third panel\footnote{\cirxone\ is
  classified as a \lmxb.  We, however, show the spectrum here since it is the
  only \lmxb\ system known that shows strong P-Cygni lines.}, however, the
  emission and absorption equivalent widths are comparable, which suggests
  that the emission component probably comes mostly from scattering of the
  continuum.  This interpretation, however, can be ambiguous since similar
  profiles can also be generated in a recombination-dominated medium with a
  small covering fraction, thereby reducing only the emission component.  The
  last panel, finally, shows a pure absorption line profile with very little
  re-emission.  In this case, the absorber, which happens to lie along the
  observers' line-of-sight, subtends only a small fraction of the sky as
  viewed from the X-ray continuum source.

\subsection{Orbital-phase Variability}

  \vela\ was observed during three different orbital phases centered on $\phi
  = 0.0, 0.25$, and $0.50$ with the \chandra\ \hetgs\ and during $\phi = 0.75$
  with \xmm.  The \hetgs\ spectra shown in Figure~\ref{fig:vela_phase} are
  easily compared with one another.  The \rgs\ spectrum is shown separately
  for clarity.

\begin{figure}[tp]
  \centerline{\psfig{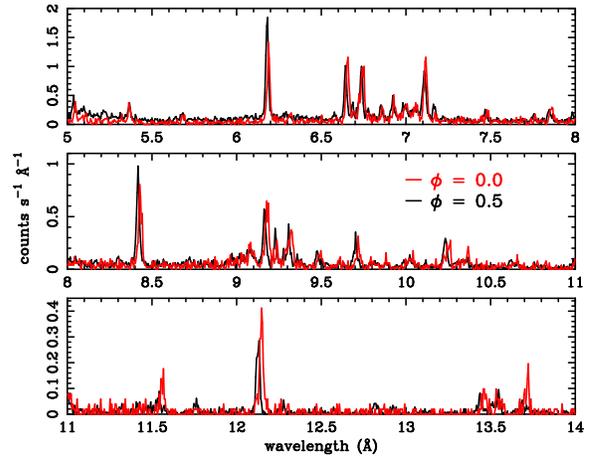}}
  \caption{Comparison of the global spectra of \vela\ at orbital phases $\phi
  = 0.0$ (red) and $\phi = 0.5$ (black).  The fluxes are rescaled to match the
  peaks of the Si~{\small XIII} forbidden lines of the two phases.  Note the
  striking similarity in the emission line intensity ratios between the two
  phases.}\label{fig:vela_var}
\end{figure}

\begin{figure}[bhtp]
  \centerline{\psfig{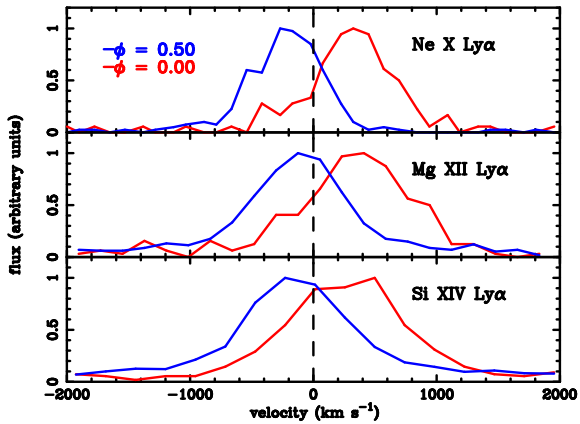}}
  \caption{Line profiles of Ly$\alpha$ transitions observed in \vela\ at two
  orbital phases; $\phi = 0.0$ (red) and $\phi = 0.5$ (blue).  The fluxes of
  each line pairs are rescaled to match at their peaks.  The width of the
  lines are due mostly to the instrument response.}\label{fig:vela_lya}
\end{figure}

  A comparison of the observed line spectra at $\phi = 0.0$ and $0.5$ is
  particularly interesting.  Shown in Figure~\ref{fig:vela_var} are the
  spectra normalized to the peak of the Si~{\small XIII} forbidden line and
  overlaid on top of each other.  Apart from the velocity shifts (see
  Figure~\ref{fig:vela_lya}), the spectra look nearly identical with $\la
  20$\% difference in the fluxes of most of the lines.  If anything, the $\phi
  = 0.5$ spectrum appears more highly-ionized as the Si~{\small XIV}
  Ly$\alpha$ line is stronger and the Si near-neutral line complex is weaker.
  This is in direct contradiction to the spherically-symmetric wind model
  presented in Sako \etal\ (1999), who predicted that the differential
  emission measure distribution at $\phi = 0.5$ would be much softer (i.e.,
  lower ionization) than during eclipse due to photoionization of more dense
  material near the stellar photosphere.  The \chandra\ data instead show that
  there is a large difference only in the overall {\emph{normalization}} of
  the distribution by roughly an order of magnitude with very little
  difference in the actual {\emph shape}.  However, this is not a surprising
  result, as disruption of the stellar wind is expected to be most significant
  near the compact object, where both the gravitational and X-ray radiation
  field are high.  A simple isolated wind model, therefore, are not applicable
  in these regions.

  Geometrically, the fact that the emission line ratios at $\phi = 0.0$ and
  $0.5$ are similar implies that the spectrum emitted in the region blocked by
  the companion during eclipse (a cylinder roughly along the line of centers)
  is nearly identical to that of the rest of the wind.  A smooth wind model
  characterized by a \cak\ velocity profile is ruled out for reasons stated
  above.  Several simple models for the clump distribution can be ruled out
  immediately as well.  First of all, a population of identical clumps
  distributed uniformly throughout the wind cannot explain this behavior,
  since clumps near the compact object will be more highly ionized.  In
  addition, a distribution of clumps, whose densities are proportional to the
  surrounding local wind density (i.e., higher density clumps closer to the
  stellar surface; see Equ.~\ref{eq:density}), cannot reproduce the data
  either.  In this case, we would expect to detect higher line fluxes from
  lower ionization species during $\phi = 0.5$, contrary to what is observed.

\begin{figure}[bt]
  \centerline{\psfig{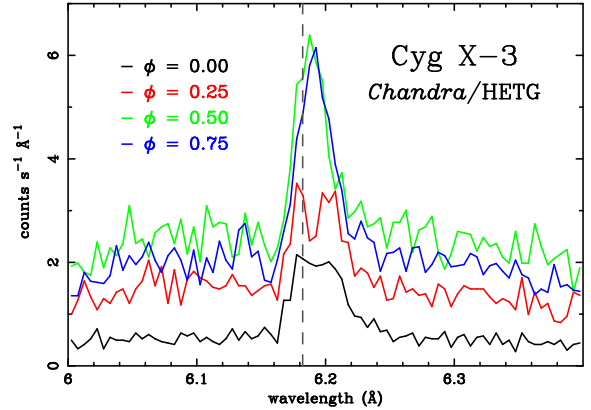}}
  \caption{Si {\small XIV} line profiles at observed at four different orbital
  phase ranges in \cygxthree.  A particularly interesting feature shown here
  is the difference in the profiles between phases $\phi = 0.75$, which shows
  a broad, smooth emission peak, and $\phi = 0.25$ that shows what appears to
  be a centrally-reversed line.  These differences might be due to radiative
  transfer effects and/or geometric occultations by, for example, the
  companion star.  In either case, the wind must be asymmetric with respect to
  the line of centers.}\label{fig:cygx3_sixiv}
\end{figure}

\begin{figure*}[ht]
  \centerline{\psfig{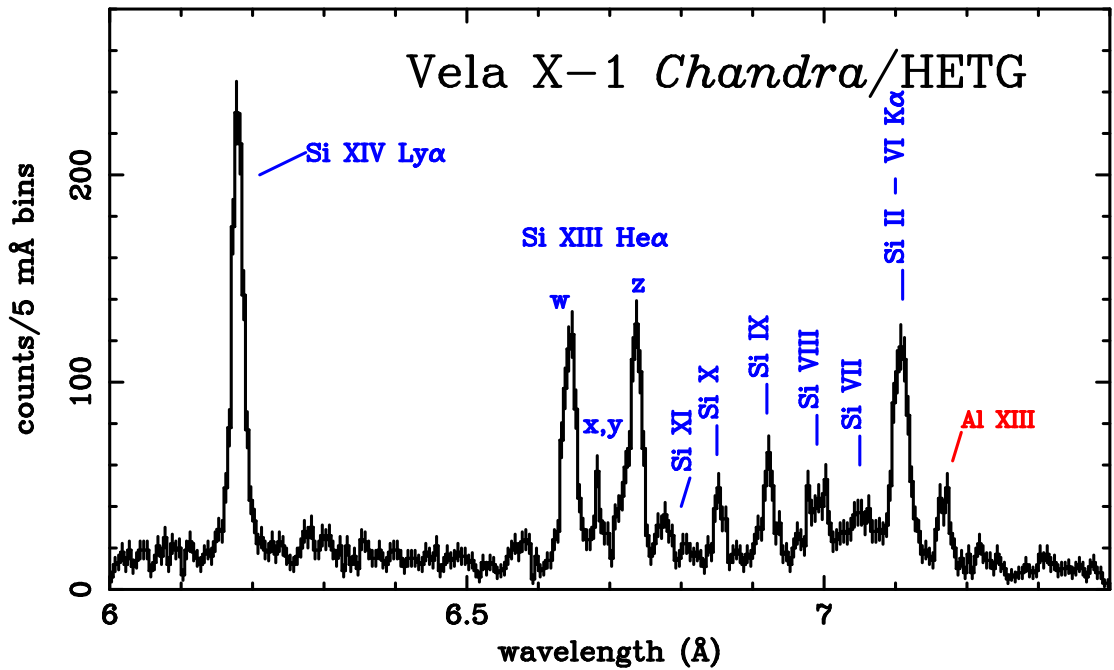}}
  \caption{The forest of Si K lines observed in the spectrum of \vela.
  K$\alpha$ lines from at least 8 charge states are detected in this narrow
  wavelength range.}\label{fig:vela_sik}
\end{figure*}

  Assuming that the wind in the shadow cone does not emit a significant amount
  of X-ray line radiation, one expects that most of the wind as seen at $\phi
  = 0.0$ would be moving away from the observer emitting lines shifted towards
  the red.  Similarly, then, lines observed during $\phi = 0.5$ should be
  blueshifted.  Qualitatively this appears to be consistent with what is
  observed in \vela\ as shown in Figure~\ref{fig:vela_lya}.  Quantitatively,
  however, they are inconsistent --- the measured line shifts and widths are
  too small.  First of all, the average line shifts of $|v| \sim 300 -
  400~\rm{km~s}^{-1}$ correspond to wind velocities at approximately $\la
  20$\% of the stellar radius from the photosphere, assuming a \cak\ profile
  given by Equ.~\ref{eq:cak} and ignoring projection effects.  Second, the
  lines have widths of $\sigma \la 200~\rm{km~s}^{-1}$ and are only marginally
  resolved with the \chandra\ \heg.  Finally, since most of the line photons
  observed during $\phi = 0.5$ come from a cylindrical region along the line
  of centers, the magnitude of the observed velocity shifts should have been
  larger than that at $\phi = 0.0$, because the wind material in this cylinder
  is moving right towards the observer at $\phi = 0.5$.  In other words, the
  observed projected velocity of $v \sim -200~\rm{km~s}^{-1}$ is probably
  close to the true wind velocity of this material.  One possibility is that
  X-ray photoionization inhibits \uv\ driving of the wind and reduces the
  terminal velocity, in which case the wind is accelerated only up a certain
  velocity and coasts at a constant velocity (Hatchett \& McCray 1977).
  Another possibility is that most of the emission does, in fact, come from a
  narrow radius interval near the surface of the companion.  In either case,
  the similarity in the line ratios between the two phases is very difficult
  to understand.

  Similar puzzling behavior has also been observed in a triple star system
  $\delta$~Ori~A (O9.5~II + B0.5~III + early B; Miller \etal\ 2002) as well as
  in the isolated B0.2 star $\tau$~Sco (Cohen \etal\ 2003), which show X-ray
  line broadening only on the order of $\sigma \la 400~\rm{km~s}^{-1}$.  These
  are again a few times smaller than the terminal wind velocity, and are also
  smaller than those seen, for example, in a younger O4f star $\zeta$~Pup
  (Kahn \etal\ 2001; Cassinelli \etal\ 2001).

  Miller \etal\ (2002) also report evidences against a simple
  spherically-symmetric wind through absorption line measurements in \cygxone.
  Using several orbital-phase resolved spectra of the source obtained with
  \chandra, they were able to qualitatively infer the mass distribution in the
  system that led them to conclude that the wind is focused along the line of
  centers, most likely through gravitational attraction by the compact object.

\section{Fluorescence Line Spectroscopy}

  Another very important discovery that resulted from the availability of high
  resolution data is the detection and identification of K-shell silicon
  fluorescent lines from a wide range of charge states.  Shown in
  Figure~\ref{fig:vela_sik} is the \hetg\ spectrum of \vela\ observed during
  $\phi = 0.5$.  K-shell fluorescent lines from essentially all charge states
  are detected in this narrow wavelength range, which covers only a few
  resolution elements of a \ccd\ spectrum.  This forest of lines would appear
  as one narrow (Si~{\small XIV}) and one broad feature (Si~{\small II} -
  {\small XIII}) as seen, for example, in the \asca\ spectrum of \vela\ (see
  Sako \etal\ 1999).

  Atomic calculations show that the strong line at $\lambda = 7.12$ \AA\ is an
  unresolved blend of K$\alpha$ lines from Si~{\small II} to Si~{\small VI}
  (fluorine-like).  The first few charge states (Si~{\small II} and {\small
  III}), however, probably do not exist in the wind of \vela\ due to the
  presence of a strong ionizing \uv\ field from the companion star.  From
  Si~{\small VII} (oxygen-like) and higher, the K$\alpha$ line complex shifts
  by $\sim 70$ m\AA\ per charge state until Si~{\small XII} (lithium-like),
  and, therefore, are well-resolved with the \chandra\ spectrometers.

  Although detailed modeling of the \vela\ spectrum is still underway (Liedahl
  \etal\ 2003), it is already clear that this provides an extremely powerful
  spectroscopic tool for studying the structure of stellar winds.  Since
  essentially all of the possible charge states are observed, the sum of the
  ion column densities yield the {\emph absolute total} column density through
  the wind, which is sensitive to the total mass loss rate of the companion
  star.  The column density distribution and the variability with orbital
  phase also allows us to infer the density spectrum and the clumping
  properties of the wind.

\begin{figure}[t]
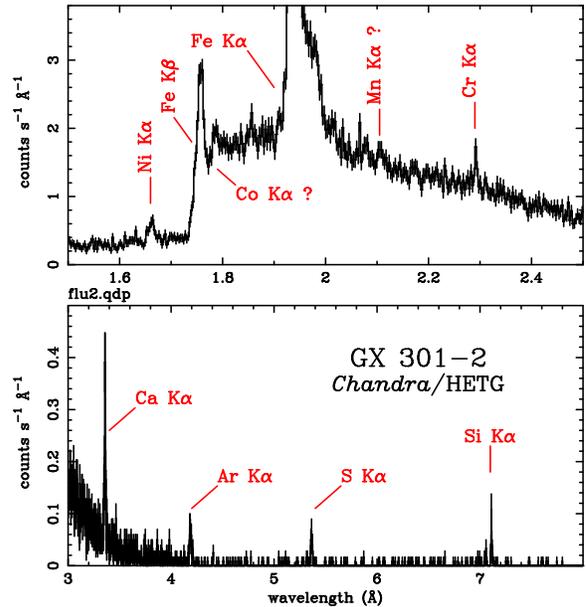

  \centerline{\psfig{file=f10a.ps,width=3.0in,angle=-90}}
  \centerline{\psfig{file=f10b.ps,width=3.0in,angle=-90}}
  \caption{Reflection-dominated spectrum of \gx\ observed during the {\small
  PP} flare.  The spectrum is highly absorbed and exhibits near-neutral
  fluorescence lines from at least 7 (possibly 9) elements.}\label{fig:gx_flu}
\end{figure}

  Finally, we note that accurate elemental abundance measurements are also
  possible with a well-exposed high-resolution spectrum.  As shown in
  Figure~\ref{fig:gx_flu}, for example, emission lines produced under similar
  physical conditions (neutral fluorescence, in this case) are detected from
  multiple elements.  Although the properties of the foreground as well as the
  embedded absorber may be rather complex, abundance ratios of elements with
  similar atomic number can nevertheless be constrained to within reasonable
  accuracy ($\la 50$\%).

\section{Compton-scattered Iron Line}

\begin{figure}[ht]
  \centerline{\psfig{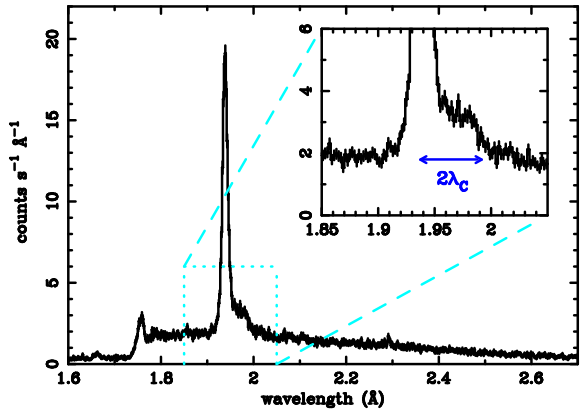}}
  \caption{The iron K line observed in the spectrum of \gx.  The K$\alpha$
  line clearly shows a broad tail towards longer wavelengths with a width of
  approximately twice the Compton wavelength (see inset).}\label{fig:gx_fek}
\end{figure}

  The spectrum obtained during the {\small PP} observation of \gx\ shows a
  spectacular broad feature that extends towards the redward side of the iron
  \kalpha line, which is identified as a Compton-scattered iron line in a low
  temperature medium.  Compton-scattered iron lines have also been observed in
  the spectra of several Seyfert galaxies (references) but with much lower
  statistical significance.  The spectrum of \gx\ shown in
  Figure~\ref{fig:gx_fek} contains $\sim 6500$ iron K$\alpha$ counts in the
  narrow component and $\sim 2000$ counts in the Compton shoulder.

  The shape and flux of the Compton profile is sensitive to the electron
  column density and temperature of the scattering medium.  The profile
  exhibits a rather sharp drop at $2\lambda_C$ from the center of the narrow
  line, which implies that the temperature cannot be much higher than $\sim
  5~\rm{eV}$.

  A detailed analysis of the spectrum and its temporal variation is presented
  in Watanabe \etal\ (2003) using detailed Monte Carlo simulations of the line
  profile.  It is found that the K$\alpha$ lines are scattered in the same
  medium where they are produced.  The similarity in the column density
  inferred from both continuum absorption and the Compton profile suggests
  that this medium surrounds the X-ray source.  The covering fraction of this
  medium with respect to the neutron star is estimated to be fairly large
  ($\Delta \Omega/4\pi \la 1$) from the measured equivalent widths of the
  emission line and the depth of the corresponding K-shell absorption edge.

  Compton broadening can, of course, occur not only in emission lines, but in
  absorption lines as well.  As in the emission line case, an intrinsically
  narrow absorption line produced in a Compton-thick medium will exhibit a
  broad red wing down to roughly twice the Compton wavelength.  This effect
  might be relevant for understanding the absorption lines observed in
  \cygxone, which show asymmetric wings towards the longer wavelengths (see,
  bottom panel of Figure~\ref{fig:p-cygni}).

\section{Future Work}

  High resolution spectroscopic data obtained with the grating spectrometers
  onboard \chandra\ and \xmm\ have provided a dynamical view of the ionized
  stellar wind in \hmxb s.  It is clear, however, that much work remains to be
  done.  In particular, detailed physical modeling of each of the individual
  sources is likely required to understand the wide variety of phenomenology
  observed in the data.  Radiative transfer effects coupled with the wind
  dynamics may be relevant for interpreting the orbital phase variability of
  the line spectra seen in some sources.  Other purely spectroscopic issues,
  such as the properties of the Si K fluorescent lines and the Compton recoil
  spectrum, must be studied in more detail to fully exploit the diagnostic
  capabilities of high resolution X-ray spectroscopy.  Despite the different
  excitation mechanisms that dominate in \hmxb\ systems and isolated OB stars,
  comparative studies of their X-ray spectra, particularly between similar
  spectral types, will be helpful for understanding the wind dynamics and how
  they are affected by the presence of the orbiting X-ray source.  Although
  additional data will also certainly be helpful for understanding some of the
  questions raised with the present data, exploratory modeling will likely
  provide new insight into the physical nature and geometry of the
  circumsource media in \hmxb\ systems.

%do not change this

%do not change this
\normalsize

\section*{ACKNOWLEDGEMENTS}

  {\small MS} was supported by \nasa\ through \chandra\ Postdoctoral
  Fellowship Award Number {\small PF}1-20016 issued by the \chandra\ X-ray
  Observatory Center, which is operated by the Smithsonian Astrophysical
  Observatory for and behalf of \nasa\ under contract {\small NAS}8-39073.
  Work at {\small LLNL} was performed under the auspices of the
  U. S. Deparment of Energy by the University of California Lawrence Livermore
  National Laboratory under contract No. W-7405-Eng-48.  {\small SW} is
  grateful for the support by research fellowships of the Japan Society for
  the Promotion of Science for young Scientists.

\end{document}